\documentclass[12pt]{article}  
\usepackage{graphicx}

\def\R{ {\rm R \kern -.31cm I \kern .15cm}}
\def\C{ {\rm C \kern -.15cm \vrule width.5pt \kern .12cm}}
\def\Z{ {\rm Z \kern -.27cm \angle \kern .02cm}}
\def\N{ {\rm N \kern -.26cm \vrule width.4pt \kern .10cm}}
\def\1{{\rm 1\mskip-4.5mu l} }
\def\lsim{\raise0.3ex\hbox{$<$\kern-0.75em\raise-1.1ex\hbox{$\sim$}}}
\def\gsim{\raise0.3ex\hbox{$>$\kern-0.75em\raise-1.1ex\hbox{$\sim$}}}
\def\noi{\noindent}

\def\beq{\begin{equation}}   \def\eeq{\end{equation}}
\def\bea{\begin{eqnarray}}  \def\eea{\end{eqnarray}}

\def\noi{\noindent}
\def\beeq{\begin{eqnarray}} \def\eeeq{\end{eqnarray}}

\begin{document}

\begin{center} {\large \bf } \vskip 2 truemm

 {\large \bf Positivity of Some Integral Transforms, and Generalization} \vskip 2 truemm
 {\large \bf of Bochner's Theorem on Functions of Positive Type}

 \par \vskip 5 truemm

{\bf Khosrow Chadan }\\ {\it Laboratoire de Physique
Th\'eorique}\footnote{Unit\'e Mixte de Recherche UMR 8627 - CNRS}\\   
{\it Universit\'e de Paris XI, B\^atiment 210, 91405 Orsay Cedex,
France}\\
{\it  (Khosrow.Chadan@th.u-psud.fr)} \par \vskip 5 truemm
\end{center}
\vskip 1 truecm

\begin{abstract}
Using the integral representations of the solutions of Schr\"odinger equation, which are the essential ingredients of the Gel'fand-Levitan and Marchenko integral equations of inverse scattering theory, we obtain a general theorem on the positivity of some integral transforms, and extend the theorem of Bochner on Fourier transforms of functions of positive type to more general transforms. The present study is restricted to the positive half-axis. We then obtain a theorem on the positivity of Fourier cosine transform of the phase-shifts.
\end{abstract}

\vskip 3 truecm 

\begin{flushleft}
LPT Orsay  07-35\\
June 2007
\end{flushleft}
 
\newpage
\pagestyle{plain}
\baselineskip 20pt
\noi {\bf I - \underline{ Introduction}} \\

In a recent paper \cite{1r}, we generalized the following theorem (Titchmarsh \cite{2r})~:\\

\noi \underline{\bf Theorem 1.} In order for $\widetilde{f}(k)$, the Fourier sine transform of $f(r)$, 
\beq
\label{1e}
\widetilde{f}(k) = \int_0^{\infty} f(r) \sin kr\ dr \ ,
\eeq

\noi to be positive, it is sufficient to have
\beq
\label{2e}
\left \{ \begin{array}{l} \hbox{$f(r)$ non-increasing over $(0, \infty )$,} \\ \\  \hbox{integrable over $(0, 1)$,} \\ \\ \hbox{and $f(r) \to 0$ as $r \to \infty$.}\end{array} \right .
\eeq

\noi It is to be remarked that $\sin kr$ is a solution of $\varphi_0'' + k^2 \varphi_0 = 0$, with $\varphi_0 (0) = 0$.\par

Consider now the (Schr\"odinger) equation 
\beq
\label{3e}
\left \{ \begin{array}{l} \varphi '' (k, r) + k^2\varphi (k, r) = V(r) \ \varphi (k, r)\ , \\ \\ r \in [0, \infty ) \ , \ V(r) > 0\ , \\ \\ \varphi (k, 0) = 0\ , \ \varphi ' (k, 0) = 1\ ,\end{array} \right .
\eeq

\noi where prime denotes differentiation with respect to $r$. The potential (positive) is supposed to satisfy the usual Bargmann-Jost-Kohn condition \cite{3r, 4r}
\beq
\label{4e}
\int_0^{\infty} r \ V(r) \ dr < \infty \ .
\eeq

\noi Under this condition, one can show that $\varphi (k, r)$ is well-defined and unique, and is, for each fixed $r$, an even entire function of exponential type in $k$, having the asymptotic behaviour \cite{3r, 4r, 5r}
\beq
\label{5e}
\varphi (k, r) =  {\sin kr \over k} + \cdots\ , \ |k| \to \infty \ ,
\eeq

\noi Also, for $k$ fixed and real, one has \cite{3r,4r,5r}
\beq
\label{6e}
\varphi (k, r) = A(k) \ {\sin (kr + \delta (k)) \over k} + \cdots\ , \ r \to \infty \ ,
\eeq

\noi where $A(k)$ is a positive factor, and $\delta (k)$ (the phase-shift) is a real and odd function of $k$. In general, for $V(r) > 0$, one has $\delta (k) < 0$, and for $V(r) < 0$, $\delta (k) > 0$ \cite{6r}.

We can now define, with the help of $\varphi (k, r)$, the integral transform of a function $f(r)$ by~:
\beq
\label{7e}
\widetilde{f}(k) = \int_0^{\infty} f(r)\  \varphi (k, r) \ dr \ .
\eeq

\noi This is a generalization of (\ref{1e}) since, for $V(r) = 0$, $\varphi$ become ${\sin kr \over k}$. Since the asymptotic behaviours of $\varphi$ for $r \to \infty$ and $k \to \infty$ are very similar to that for $V(r) = 0$, one can use all the machinery of usual Fourier integrals to study the convergence, summability, and inverse transforms of (\ref{7e}) \cite{2r,7r}. In \cite{1r}, we generalized Theorem 1 to~: \\

\noi \underline{\bf Theorem 2.} In order for $\widetilde{f}(k)$, defined by (\ref{7e}), to be positive, it is sufficient for $f(r)$, to be of the form 
\beq
\label{8e}
f(r) = \int_r^{\infty} \left [ \chi_0 (r) \ \varphi_0(t) - \varphi_0 (r) \ \chi_0 (t) \right ] g(t)\ dt \ ,
\eeq

\noi where $g(t)$ is an arbitrary positive function such that the integral converges at infinity. Here, $\varphi_0 (r) \equiv \varphi (k=0, r)$ is the solution of (\ref{3e}) at zero energy, and $\chi_0 (r)$ is the second, independent solution at $k= 0$, defined by 
\beq
\label{9e}
\chi_0 (0) = 1\ ,\ W(\varphi_0, \chi_0) \equiv \varphi '_0 \chi_0 - \varphi_0 \chi '_0 = 1 \ .
\eeq

It is well-known that since $V(r) > 0$, $\varphi_0 (r)$ is an increasing convex function of $r$, and one has \cite{3r,4r,5r}~:
\beq
\label{10e}
\left \{ \begin{array}{l} \varphi_0 (r) \ \displaystyle{\mathrel{\mathop =_{r \to \infty}}}\  A_0 r + B_0 + \cdots \ ,\\ \\ A_0 > 1\quad , \quad B_0 < 0 \end{array}\right .
\eeq

\noi where $A_0$ is given by $A(k=0)$ of (\ref{6e}). Remember that, by definition, $\varphi_0(0) = 0$, and $\varphi '_0(0) = 1$. It can be checked now in a straightforward manner that one can define $\chi_0 (r)$ by
\beq
\label{11e}
\chi_0 (r) = \varphi_0 (r) \int_r^{\infty} {dt \over \varphi_0^2 (t)} \ dt \ .
\eeq

It is then easily seen that, indeed,
\beq
\label{12e}
\chi_0 (0) = 1 \ , \ \lim_{r\to 0}\  r \ \chi_0 (r) = 1\ , \ \chi_0 (\infty ) = {1 \over A_0} \ .
\eeq

It follows that $\chi_0 (r)$ is a decreasing convex function of $r$. From (\ref{11e}), formula (\ref{8e}) cna also be written as
\beq
\label{13e}
f(r) = \varphi_0 (r) \int_r^{\infty} \varphi_0 (t)\ g(t) \ dt \int_r^t {du \over \varphi_0^2 (u)}\ .
\eeq

\noi In this formula, all the terms are positive, and so it is all easy to find what one has to impose on $g(t)$ in order to have $f(r) \in L^p (0)$, or $L^p (\infty )$, ... etc. \par

One finds that
\beq
\label{14e}
\left \{ \begin{array}{l} h_1(r) \equiv \displaystyle{\int_r^1} tg (t) \in L^p (0, 1) \Leftrightarrow f(r) \in L^p (0, 1) \ ,\\ \\ h_2(r) \equiv \displaystyle{\int_r^{\infty}} tg(t) \ dt \in L^p (1, \infty) \Leftrightarrow f(r) \in L^p (1, \infty) \ . \end{array}\right .
\eeq
\vskip 5 truemm

\noi {\bf Remark.} For having $f(r) \in L^1 (0, 1)$ as in {\bf  Theorem 1} of Titchmarsh, it is sufficient for $g(r)$ to be less singular than $r^{-3}$ at the origin. Also, in order to have $f(\infty ) = 0$, it is sufficient to have $rg \in L^1 (1, \infty )$. The only difference between our conditions on $f$ in {\bf Theorem 2}, and the conditions in {\bf Theorem 1} is that now, as can be seen from (\ref{13e}), we have 
\beq
\label{15e}
f'' = V(r) f + g(r)\ , 
\eeq

\noi which means that, even when $V(r) = 0$, we must have $f(r)$ convex, whereas, in {\bf Theorem 1}, $f(r)$ had to be only non-increasing. The reason is that our Theorem~2 is quite general, and applies with any positive potential $V(r)$ in (\ref{3e}) and (\ref{4e}), whether monotonous (decreasing !) or not. If we are willing to assume that $V(r)$ is non-increasing, then it can be shown that we have :\\

\noi \underline{\bf Theorem 1${\bf '}$.} If $V(r)$ is non-increasing, then Theorem 1 applies as well to (\ref{7e}) under the same conditions on $f(r)$, i.e. $\widetilde{f}(k)$ is positive if one has (\ref{2e}). The proof mimiks the simple proof of Theorem 1, as given in \cite{2r}. Taking, as example, the centrifugal potential $\ell (\ell + 1)/r^2$, $\ell > 0$, one is led to Hankel transforms where one finds many examples of integrals in which $f(r)$ satisfies (\ref{2e}), and which are positive [2, chap. 8].\newpage

\noi {\bf II. \underline{Bochner's Theorem}}\\

The purpose of the present paper is to generalize, using the same technique as in \cite{1r}, the following theorem \cite{7r,8r,9r}~:\\

\noi \underline{\bf Theorem 3 (Bochner)} : If $\alpha (t)$ is a non-decreasing bounded function on $(- \infty , \infty )$, and if $F(x)$ is defined by the Stieltjes integral
\beq
\label{16e}
F(x) = \int_{-\infty}^{\infty} e^{ixt} \ d \alpha (t) \ , \quad - \infty < x < \infty \ , 
\eeq

\noi then $F(x)$ is a continuous function of positive type. We recall the reader that a (not necessarily measurable) function $F(x)$ defined on $(- \infty , \infty )$ is said to be of positive type if
\beq
\label{17e}
\sum_{m=1}^s \ \sum_{n=1}^s a_m\ \overline{a}_n \ F(x_m - x_n) > 0
\eeq

\noi for any finite number of arbitrary real $x_1, \cdots , x_s$ and a like number of complex $a_1, \cdots , a_s$. Conversely, if $F(x)$ is measurable on $(- \infty , \infty )$, and $F$ is of positive type, then there exists a non-decreasing bounded function $\alpha (t)$ such that $F(x)$ is given by (\ref{16e}) for almost all $x$, $- \infty < x < \infty$. We should remark that, in the converse theorem, Bochner assumed $F(x)$ to be continuous, and showed that $\alpha (t)$ is such that (\ref{16e}) is true for all $x$. F. Riesz showed that measurability of $F(x)$ was sufficient in the converse theorem.\par

We are now going to generalize the above theorem by replacing, as we did similarly in \cite{1r}, the exponential function in (\ref{16e}) by the appropriate solution of the differential equation (\ref{3e}), namely the Jost solution, which satisfies \cite{3r, 4r, 5r}
\beq
\label{18e}
\left \{ \begin{array}{l} f''(k,r) + k^2 f(k,r) = V(r) \ f(k,r) \ , \\ \\ r\in [0, \infty ) \ , \ V(r) > 0\ , \ V \in L^1(0,1)\ , \ r V\in L^1 (1 , \infty )\ ,\\ \\ \displaystyle{\lim_{r \to \infty}}\ e^{-ikr} \ f(k, r) = 1\ , \\ \\  \displaystyle{\lim_{r \to \infty}}\ e^{-ikr} f'(k,r) = ik \ . \end{array}\right .
\eeq

\noi Moreover, for each fixed value of $r$ ($\geq 0$), $f(k, r)$ is holomorphic and bounded for $k$ in $Im\ k >0$. It vanishes exponentially there as $|k| \to \infty$. For $k$ real, $f(k,r)$ is simply bounded.\par

Here, since we consider the half-axis $r \in [0, \infty )$, we must restrict the support of $\alpha (t)$ in (\ref{16e}) to be also in $[0, \infty )$. We shall study the case of the full axis $x \in (- \infty, \infty )$ in a separate paper. \\

\noi {\bf Remark.} If the support of $\alpha (t)$ is restricted to the half axis $t \geq 0$ in (\ref{16e}), it is obvious that $F(x)$ can be extended analytically in the upper half-plane $Im\ x > 0$, and it is holomorphic and bounded there.\\

The theorem which generalizes {\bf Theorem 3} of Bochner is now as follows~: \\

\noi Consider the Stieltjes integral
\beq
\label{19e}
\widetilde{f}(k) = \int_0^{\infty} f(k, r) d\alpha (r) \ ,
\eeq

\noi where $f(k,r)$ is the (Jost) solution defined by (\ref{18e}). \\

Now the Jost solution defined by (\ref{18e}) has the integral representation [5, chap. V ; 11, chap. 4]~:
\beq
\label{20e}
f(k,r) = e^{ikr} + \int_r^{\infty} A(r, t)\  e^{ikt} dt \ ,
\eeq

\noi where the kernel $A(r, t) \in L^1 (r , \infty ) \cap L^2(r, \infty )$ in $t$, and is the solution of the integral equation
\beq
\label{21e}
A(r, t) = {1 \over 2} \int_{{r+t \over 2}}^{\infty} V(s)\ ds + \int_{{r+t \over 2}}^{\infty} ds \int_0^{{t-r \over 2}} V(s-u) \ A(s-u, s+ u) \ du\ .
\eeq

\noi It can be shown that this integral equation has a unique positive solution obtained by iteration (absolutely convergent series~!) and satisfies the bound ($V(r) > 0$~!)~: 
\beq
\label{22e}
0 < A(r, t) \leq {1 \over 2} \int_{{r+t \over 2}}^{\infty} V(s) \ ds \left [ \exp \int_r^{\infty} u V(u) \ du \right ] \ .
\eeq

\noi Note here that the integrals are absolutely convergent. Also, it is obvious on (\ref{22e}) and the integrability conditions on $V(r)$ shown in (\ref{18e}), that $A(r, t)$ is a bounded continuous function, and goes to zero at infinity when $r (\leq t) \to \infty$ or $t \to \infty$.  We can then replace in (\ref{19e}) $f(k, r)$ by its integral representation (\ref{20e}), exchange the order of integrations, to find \cite{2r,7r,8r}
$$\hskip 4 truecm \left \{ \begin{array}{l} \widetilde{f}(k) = \displaystyle{\int_0^{\infty}} e^{ikr}\ d \beta (r) \ , \hskip 5.5 truecm (23{\rm a}) \\ \\ d\beta (r) = d \alpha (r) + \left ( \displaystyle{\int_0^r} A(t, r) d \alpha (t) \right ) dr \ . \hskip 2.7 truecm (23{\rm b})
 \end{array}\right .$$

\noi $A(t, r)$ being positive, it follows that $\beta (r)$ is bounded and increasing, if $\alpha (r)$ is so. We can therefore mimic the proof of theorem 3, as given in \cite{8r}, and prove the existence of $\beta (r)$, bounded and increasing, by using the second part of Theorem 3. Once the existence of $\beta (r)$ is shown, one has to solve the Volterra integral equation \cite{12r}
$$\alpha '(r) = \beta ' (r) - \int_0^r A(t, r) \ \alpha ' (t)\ dt \ . \eqno(24)$$

\noi The kernel $A(t,r)$ being a bounded continuous function, it is known that (24) has a unique solution obtained by iteration, i.e. iterating (24), by starting from $\beta '(r)$, we obtain an absolutely and uniformly convergent series defining the solution \cite{12r}. Moreover, since $A(t, r) \to 0$ as $r \to \infty$, all the higher terms of the series go to zero. In fact, they are all $L^1 (\infty )$ since
$$\int_r^{\infty} V(s) \ ds \in L^1 (1, \infty )\ . \eqno(25)$$

\noi The first term of the series, namely $\beta '(r)$ being itself $L^1$, it is obvious that the solution $\alpha '(r)$, given by an absolutely and uniformly convergent series, is also $L^1$. If we call $M$ the absolute bound of $A(r, t)$ for all $0 \leq r \leq t \leq \infty$, it is trivial to find
$$|\alpha '(r) | < \beta ' (r) \ e^{M\beta (r)}\ . \eqno(26)$$

\noi Since $\beta (r)$ is a bounded increasing function of $r$, it follows that $\alpha (r)$ is a bounded function. Putting together everything, we have~:\\

\noi \underline{\bf Theorem 4.} Consider the Stieltjes integral (\ref{19e}), with $\alpha (r)$ positive, bounded, and non-decreasing. Then $\widetilde{f}(k)$ is a function of positive type having the usual representation (23a, b). Conversely, if we consider a function $\widetilde{f}(k)$, holomorphic and bounded in $Im\ k > 0$, and of positive type, then it can be represented in the form (\ref{19e}), where $\alpha (r)$, bounded, but not necessarily positive or non-decreasing, is given by the unique solution of the Volterra integral equation (24). The kernel $A(t, r)$ itself is defined by the unique solution of the integral equation (\ref{21e}), $V(r)$ being the potential defining the integral representation (\ref{19e}) via (\ref{18e}).\\

\noi {\bf Remark.} Notice the unsymmetry between $\alpha (r)$ and $\beta (r)$. If $\alpha (r)$ is non-decreasing, so is also $\beta (r)$. However, the converse is not true, as seen on (24), unless $A(t,r)$ is small, so that, in the iteration of (24), the dominant term is $\beta'(r)$. And for $A(t, r)$ to be small enough, one sees on (\ref{22e}) that $V(r)$, positive, must be small enough.\\

\noi {\bf III. \underline{Applications}}\\

In ref. \cite{1r}, we gave an application of our theorem 2 to secure the absence of positive energy bound states (bound states embedded in the continuum) in the radial Schr\"odinger equation for a class of nonlocal potentials. We give now an application of the Bochner's Theorem to the Fourier integral representation of the phase-shift. \par

We consider again the $S$-wave for simplicity, i.e. equation (\ref{3e}) with condition (\ref{4e}). It can be shown that the phase shift $\delta (k)$ can be written as \cite{5r}
$$ \left \{ \begin{array}{l} \delta (k) = - \displaystyle{\int_0^{\infty}} \gamma (t) \sin kt\ dt\ , \\  \\ \gamma (t) \ \hbox{real and $\in L^1(0, \infty )$}\ .   \end{array}\right . \eqno(27)$$

\noi The minus sign in front of the integral is just for convenience. Clearly, $\delta (k)$ is a continuous and bounded function of $k$ for all $k \geq 0$, and vanishes at $k = 0$ and $k = \infty$. Moreover, one can show that \cite{5r}
$${\delta (k) \over k} \ \in L^1(0, \infty )\ . \eqno(28)$$

Another representation of the phase-shift is the following \cite{14r} 
$$\delta (k) = - k \int_0^{\infty} V(r)\ {\varphi^2 (k, r) \over \varphi{'}^2 (k, r) + k^2 \varphi^2 (k, r)}\ dr\ , \eqno(29)$$

\noi where $\varphi (k, r)$ is the physical solution defined in (\ref{3e}), the integral being absolutely convergent.\\

\noi {\bf Remark.} For $k$ real and $\not= 0$, the fraction under the integral 
$$\phi (k, r) = {\varphi^2 (k, r) \over \varphi{'}^2 (k, r) + k^2 \varphi^2 (k, r)} \eqno(30)$$

\noi is always a bounded function for all $k \geq 0$, and all $r \geq 0$. Indeed, for $k > 0$, $\varphi$ and $\varphi '$ canot both vanish simultaneously for some $r = r_0$ without having $\varphi \equiv 0$ \cite{10r}. For $r = 0$, the denominator is just 1, by definition. For $k = 0$, the denominator reduces to $\varphi{'}^2(0, r)$, and because of $V(r) \geq 0$, $\varphi ' (0, r)$ is an increasing function of $r$, starting from $\varphi '(0, r=0) = 1$ \cite{1r,10r}. Moreover, because of (\ref{5e}), we have, for each $r$ $(\geq 0)$ fixed,
$$\phi (k, r) = {\sin^2 kr \over k^2} \ + \cdots \ , \quad k \to \pm \infty \eqno(31)$$

\noi so that $\phi (k, r)\in L^1 (0, \infty )$ in the variable $k$.\par

From formula (29), one can show that $\delta (k)$ is a differentiable function of $k$ for all $k > 0$ \cite{14r}. In order tos ecure also the differentiability at $k = 0$, one needs to impose the extra condition at infinity~:
$$r^2 V(r) \in L^1(1, \infty )\ . \eqno(32)$$

\noi We can summarize the above results in the following known~:\\

\noi \underline{\bf Theorem 5.} Under the condition (\ref{4e}) on the potential, the phase-shift $\delta (k)$ is a continuous and bounded function of $k$ for all $k \geq 0$, and satisfies (28). It is also continuously differentiable for all $k > 0$. If (32) is also satisfied, the derivative exists for $k = 0$, and is finite. Obviously, $\delta (0) = \delta (\infty ) = 0$.\\

We introduce now
$$\Gamma (t) = \int_t^{\infty} \gamma (u)\ du \ . \eqno(33)$$

\noi By definition, $\Gamma (t)$ is a bound and continuous function of $t$ for all $t \geq 0$, and $\Gamma (\infty ) = 0$. Using now $\gamma (t) = - \Gamma '(t)$ in (27), and integrating by parts, we find
$$\delta (k) = - k \int_0^{\infty} \Gamma (t)\cos kt\ dk \ , \eqno(34)$$

\noi the integral being convergent at infinity by the Abel lemma \cite{2r}. Comparing (34) with (29), and inverting the Fourier cosine transform, we get
$$\Gamma (t) = {2 \over \pi} \int_0^{\infty} {- \delta (k) \over k} \cos kt\ dk = \int_0^{\infty} \left [ \int_0^{\infty}  V(r) \  {\varphi^2 \over \varphi{'}^2 + k^2 \varphi^2}\ dr \right ] \cos kt\ dk$$
$$= \int_0^{\infty} V(r) dr \int_0^{\infty} {\varphi^2 (k, r)  \over \varphi{'}^2  + k^2 \varphi^2}\ \cos kt\ dk\ , \eqno(35)$$

\noi the exchange of the two integrations being allowed by virtue of the remark after (29), i.e. (31). \par

For each fixed $r \geq 0$ $\phi (k, r)$, defined by (30), is a real, bounded, and continuous function of $k$ for all $k \geq 0$, and vanishes at $k = \pm \infty$; Obviously, it is also positive, and even in $k$. Therefore, it is straightforward to show that $\phi$ is a function of positive type as was defined in Theorem 3. It follows that, according to the Theorem, for each $r$ fixed $(\geq 0$), we have
$$\phi (k, r) = \int_{-\infty}^{\infty} e^{ikt}\ d \alpha (r, t)\ , \eqno(36)$$

\noi $\alpha$ being a bounded non-decreasing function of $t$. However, $\phi (k, r)$, satisfying (31), is $L^1(0 , \infty )$ in $k$. The inversion of (36) then shows that, in fact $\dot{\alpha} (r, t) = d\alpha (r, t)/dt$ is continuous and bounded, and vanishes at $t = \pm \infty$. Using now the fact that $\phi (k, r)$ is a real even function of $k$, we can write (36) as
$$ \left \{ \begin{array}{l} \phi (k, r) = \displaystyle{\int_0^{\infty}} \omega (r, t) \cos kt\ dt\ , \\  \\ \omega (r, t) = \displaystyle{{1 \over 2}} \left [ \displaystyle{{d \over dt}} \ \alpha (r, t) + \displaystyle{{d \over dt}} \ \alpha (r, -t)\right ] > 0\ ,   \end{array}\right . \eqno(37)$$

\noi $\omega (r, t)$ being a bounded and continuous function of $t$, and
$$\omega (r, \infty ) = 0 \ . \eqno(38)$$

\noi Since $\phi (k , r)$ is $L^1$ in $k$, we can invert (37), and write
$$\omega (r, t) = {2 \over \pi} \int_0^{\infty} \phi (k, r) \cos kt\ dk \ ,  \eqno(39)$$

\noi the integral being absolutely convergent. Therefore, $\omega (r, t)$, for each $r \geq 0$, is a continuous and bounded function of $t$ for all $t \geq 0$. This, used now in (35), leads to
$$\Gamma (t) = {2 \over \pi} \int_0^{\infty}\omega (r, t)\ V(r) \ dr > 0\ , \eqno(40)$$

\noi since both $\omega$ and $V$ are positive. We should remark here that, from the definitions (30) and (39), it follows from (\ref{3e}) that $\omega (r, t)$ is $0(r^2)$ as $r \to 0$. Also, we have (38). Therefore, the integral in (40) is absolutely convergent, and defines a bounded function of $t$ for all $t \geq 0$. We can therefore summarize our result in the following \\

\noi \underline{\bf Theorem 6.} Under the assumption (\ref{4e}) on the positive potential $V(r)$, the phase-shift has the integral representation (34), where $\Gamma (t)$, a continuous and bounded function according to its definition (33), and vanishing at $t = \infty$, is positive.\\

\noi {\bf Remark.} Formula (29) is quite general, and is valid for all $V$ satisfying $r V(r) \in L^1 (0, \infty )$, whether positive or not. It shows the well-known fact the phase-shift has the opposite sign of $V$ in cases where $V$ has a definite sign \cite{6r}. When $V(r)$ is negative, and admits some bound states of energies $- \gamma_j^2$, $j = 1, \cdots , n$, one can show that \cite{3r,5r}
$$\widetilde{\delta} (k) = \delta (k) - 2\sum_j  {\rm Arctg}\ {\gamma_j \over k} \eqno(41)$$

\noi has a similar representation as (27), and one has, of course, $\widetilde{\delta} (0) = \widetilde{\delta} (\infty ) = 0$. One may then be tempted to apply our Theorem 6 to $\widetilde{\delta} (k)$. However, it is not obvious that $\widetilde{\delta} (k)$ corresponds to a positive potential $\widetilde{V} (r)$. The corrresponding potential may be oscillating, while being weak enough not to admit bound states.\\

\noi \underline{\bf Acknowledgements.} The author would like to thank Professors Kenro Furutani and Reido and Takao Kobayashi, as well as the Science University of Tokyo (Noda), for very warm hospitality and financial support when this work began.

\end{document}